\documentclass[numbers]{article}

\usepackage[final]{neurips_2021}

\usepackage[utf8]{inputenc} % allow utf-8 input
\usepackage[T1]{fontenc}    % use 8-bit T1 fonts
\usepackage{url}            % simple URL typesetting
\usepackage{booktabs}       % professional-quality tables
\usepackage{amsfonts}       % blackboard math symbols
\usepackage{nicefrac}       % compact symbols for 1/2, etc.
\usepackage{microtype}      % microtypography
\usepackage{xcolor}         % colors
\usepackage{graphicx}
\usepackage{subcaption}
\usepackage{wrapfig}

\title{Bridging the gap to real-world for network intrusion detection systems with data-centric approach}

\author{%
  Gustavo de C. Bertoli\\
  PhD Candidate\\
  Aeronautics Institute of Technology\\
  \texttt{gustavo.bertoli@ga.ita.br} \\
  \And
  Lourenço A. Pereira Jr.,\, Filipe A. N. Verri \\
  Department of Computer Science\\
  Aeronautics Institute of Technology\\
  \texttt{ljr,verri@ita.br} \\
   \And
  Aldri Santos\\
  Department of Computer Science\\ 
  Federal University of Minas Gerais\\  
  \texttt{aldri@dcc.ufmg.br} \\ 
  \And
  Osamu Saotome\\
  Department of Electronics Engineering\\
  Aeronautics Institute of Technology\\
  \texttt{osaotome@ita.br} \\
}

\begin{document}

\maketitle

\begin{abstract}
Most research using machine learning (ML) for network intrusion detection systems (NIDS) uses well-established datasets such as KDD-CUP99, NSL-KDD, UNSW-NB15, and CICIDS-2017. In this context, the possibilities of machine learning techniques are explored, aiming for metrics improvements compared to the published baselines (model-centric approach). However, those datasets present some limitations as aging %(e.g., aging, specific network architectures) 
that make it unfeasible to transpose those ML-based solutions to real-world applications.
This paper presents a systematic data-centric approach to address the current limitations of NIDS research, specifically the datasets. This approach generates NIDS datasets composed of the most recent network traffic and attacks, with the labeling process integrated by design.
\end{abstract}

\section{Introduction}

There is an increase of connected devices nowadays as aggregated by the Internet of Things (IoT) paradigm and further with the rise of 5G~\cite{iot_forecast, ericsson_mobility_report}. This scenario results in a greater heterogeneity of devices and network architectures %(e.g., F/M/VANETs) 
to make the solutions available, such as autonomous driving, remote surgery, connected infrastructures, among others. In this context, cybersecurity is one of the properties that must be in place to make these solutions available with trust. 
%Critical applications that were previously offline are now becoming connected to the Internet, exposing them to remote threats leading to risks such as harmful events to people or society and compromising privacy~\cite{colonialpipeline, solarwinds}. %\textbf{COLOCAR MAIS UMA OU DUAS SENTENÇAS DESENVOLVENDO A SENTENÇA ANTERIOR}

To deal with the pace of increasing network agents, architectural changes, diversity of attacks, and the increasing amount of network traffic, the use of machine learning (ML) based network intrusion detection systems (NIDS) raises as a technical approach to deal with the evolving context bringing trust to these applications~\cite{anderson2020security}. 
In contrast with the NIDS research trend, ML-based NIDS are still not prevalent in real-world applications~\cite{nids2021}. Instead, most research uses well-established datasets like KDD-CUP99~\cite{kdd-cup99}, NSL-KDD~\cite{nsl-kdd}, UNSW-NB15~\cite{unsw-nb15}, and CICIDS-2017~\cite{cicids2017}, that from a machine learning perspective, are good as baseline comparisons on different ML techniques. However, to derive these findings to real-world application, we claim that a data-centric approach %in a systematic process 
must be in place to continuously generate datasets and re-train models, address the following limitations: evolving network traffic, aging of attacks, and capability to generalize for different network architectures.

The remaining of 
the paper presents the related works about the challenges faced by NIDS datasets in Section~\ref{related}. Section~\ref {methodology} and \ref{results} details our approach to generate NIDS datasets from a data-centric perspective. Section~\ref{conclusion} presents the conclusion, next steps of our research, and open questions.

\section{Challenge and Related Work}\label{related}
\cite{nids2021} reported the low performance of NIDS in the real-world environment as a research challenge. Some probable causes are using old datasets and not evaluating the proposed solutions considering a realistic environment. In addition, the authors report the lack of a systematic approach for dataset generation capable of being updated frequently.
The challenges about datasets for NIDS research are also reported by~\cite{paxson2010}. It presents the inherent network traffic diversity characteristic and the difficulty for a dataset to cope with all possibilities. Furthermore, it is highlighted the difficulty of data availability (public datasets) and criticizes generalizing results obtained in small network architecture to larger networks. Regarding dataset aging, \cite{paxson2010} reports that the most used datasets in 2010 were already one decade old (DARPA98, and KDD-CUP99). \cite{ml_survey_2019, ml_cyber} highlights the current trend for using a deep learning approach in the NIDS research but also points out the use of old datasets such as KDD-CUP99 and NSL-KDD% (based on KDD-CUP99)
. It also reports the lack of available datasets and low performance in a real-world environment. Next, \cite{taxonomy_nids_2020} provides a survey about NIDS datasets and reports the outdated datasets in use by NIDS research as a challenge.

The survey by \cite{survey_datasets_2019} reports the ``perfect dataset'' composed of up-to-date traffic, labeled, publicly available, with real network traffic and a multitude of attacks and normal traffic, spanning a long time frame. However, they are skeptical about the availability of datasets comprising all these properties, with the challenge for a labeled dataset containing long-time traffic.
In this context of dataset aging for NIDS applications, it is important to clarify that the network traffic and attacks have a lifespan. In other words, the NIDS are susceptible to concept drift as reported by~\cite{lifespan}. It reports a six-week of good performance of the trained models. This study indicates that a continuous update is required to address the evolving characteristics of the network traffic.

In summary, the related works present dataset aging and poor performance in a real-world environment as a research gap. Our proposition is a framework for dataset generation capable of generating NIDS datasets with the most recent network traffic and attack patterns shifting the current NIDS paradigm from model-centric to data-centric. The downside of our approach is the limited possibilities to benchmark ML solutions using the generated dataset with published baselines. Nevertheless, we envision this approach as a next step to the current NIDS solutions to bridge the gap to the real world. Thus, after performing the traditional NIDS research to obtain the best model, our approach must be part of an end-to-end pipeline to continuously re-train this model with a dataset composed of up-to-date traffic and network attacks. Another limitation is that we do not evaluate the use of commonality for NIDS features between our generated datasets and those publicly available~\cite{features}.

\section{Methodology}\label{methodology}

\begin{figure}[h]
    \centering
    \includegraphics[width=0.8\textwidth]{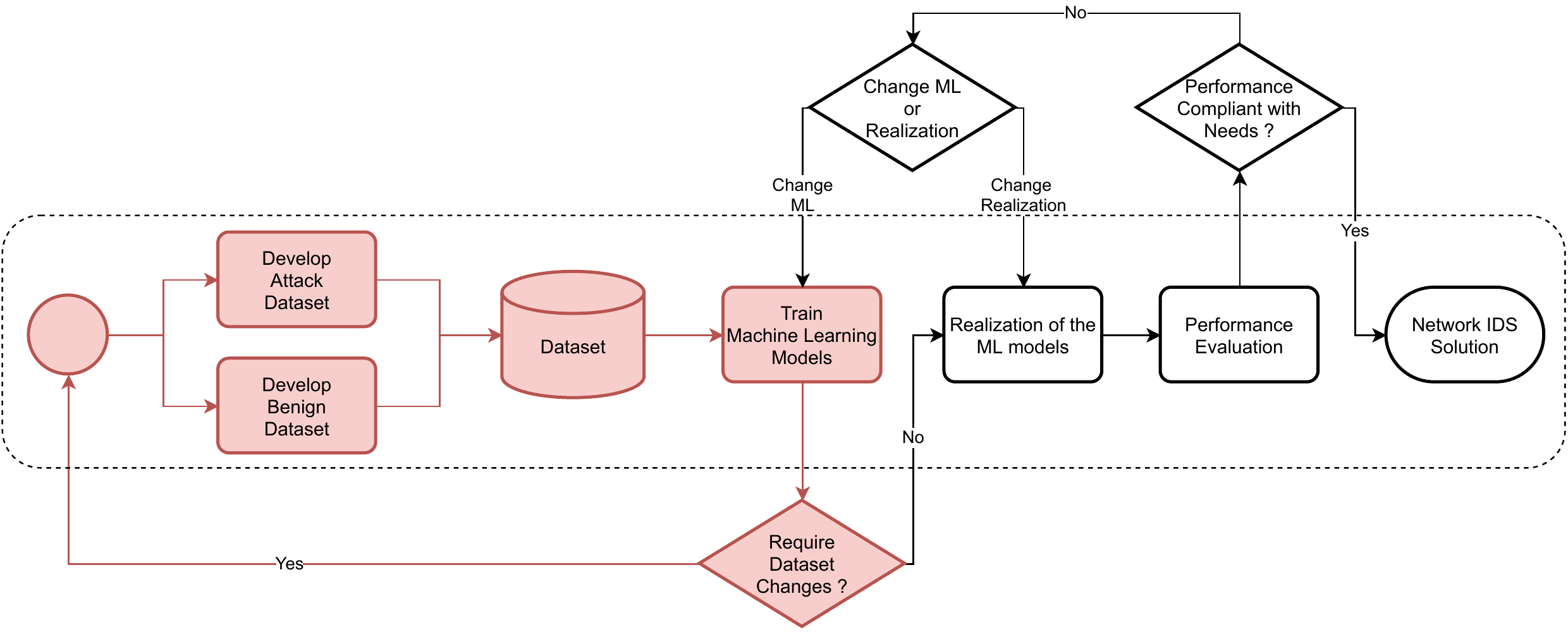}
    \caption{Workflow for the development of a ML-based NIDS solution. Dataset is part of the process.}
    \label{fig:workflow}
\end{figure}

Our methodology starts with the generation of network traffic that represents the attack behavior. This step aims to overcome the current datasets' challenges: not containing the most recent attacks, not reproducible, and not properly labeled~\cite{halfakingdown2015}. For the attack traffic, we take advantage of virtualization and cloud infrastructure to generate this traffic on-demand.
We use a Docker container with the base image of Kali Linux distribution (a well-known distribution for cybersecurity tasks). The use of Kali Linux is a common approach to generate attack traffic~\cite{ton_iot}. This container is responsible for generating all attack traffic against the cloud infrastructure that we manage. This container is a customized step that can include different attacks and targets through configuration files. Knowing the attacker instance and the targets in advance is crucial because this information supports the automation of the labeling process. For example, it is possible to construct a tuple with the attacker and target IP addresses to determine the attack traffic.% or change the container's IP address according to each specific attack for the labeling process.
To compose the NIDS dataset, we use a method known as salting~\cite{salting2011}. This method merges the legit traffic data with the attack traffic generated on-demand. We obtain this legit traffic after removing the malicious traffic from open-data traffic providers as MAWILab~\cite{mawilab}, from other public NIDS datasets, or from the deployment environment that increases the effectiveness of the solution (better data). Using traffic from the deployment environment requires authorization of the IT infrastructure in conjunction with the network administrators to label the legit traffic correctly. This labeling process can make use of firewall rules to determine the legit traffic. 
The generation of the dataset becomes part of the overall process of NIDS development (Figure~\ref{fig:workflow}). This process considers the generation of up-to-date datasets from both attack and benign traffic perspectives to bridge the gap between research and the real-world application of ML-based NIDS.

We also ensure that the data generated %in our framework 
is \emph{tidy}~\cite{wickham2017}, that is, has a rectangular structure composed of observations (rows) and variables (columns).  Such an organization, popular among data scientists, enables the utilization of common tools for exploratory analysis, data manipulation, and model building~\cite{wickham2017}. 
Each packet in the dataset has its row (Table~\ref{tab:dataset}). Each column represents a packet feature that can be either context-aware ($X^C_1$, \dots, $X^C_{n_c}$) that are common to multiple observations, such as source and destination IP addresses or context-free, that is, intrinsic of the packet ($X^P_1$, \dots, $X^P_{n_p}$).
One of the columns ($Y$) represents the variable that informs whether the packet is from benign traffic or an attack.
\begin{wraptable}{r}{0.5\textwidth}
%\begin{table}[h]
    \centering
    \caption{Shape of the data generated in our framework. Context-aware and packet-intrinsic features.}
    \label{tab:dataset}
    \begin{tabular}{c c c c c c c}
        \toprule
        $X^C_1$ & \dots & $X^C_{n_c}$ & $X^P_1$ & \dots & $X^P_{n_p}$ & $Y$ \\
        \midrule
        \dots & \dots & \dots & \dots & \dots & \dots & \dots \\
        \dots & \dots & \dots & \dots & \dots & \dots & \dots \\
        \dots & \dots & \dots & \dots & \dots & \dots & \dots \\
        \dots & \dots & \dots & \dots & \dots & \dots & \dots \\
        %\dots & \dots & \dots & \dots & \dots & \dots & \dots \\
        \bottomrule
    \end{tabular}
%\end{table}
\end{wraptable}
  
% TODO, explain that rows/observations are packets and columns/variables are context-aware features (IP, time since first packet, id of the connection, protocol, etc.) and context-free features (intrinsic properties of the packet). [Although there is an obvious functional relationship between all features]
% Nesse parágrafo acima, o ideal é trazer uma linguagem do domínio de aplicação.

As a result, one can easily reshape the generated dataset to meet the requirements and aspects of the desired machine-learning task.  For instance, in a stateless approach, one can discard all context-aware variables (Table~\ref{tab:stateless}); while, in a stateful approach, one can use the consolidated summarizing operations in data science to aggregate rows (Table~\ref{tab:stateful}).  The common and consistent shape of the data also eases selection and filtering operations.

\begin{table}[h]
  \caption{Different shapes for manipulated data.}
  \begin{subtable}[t]{0.48\columnwidth}
    \centering
    \caption{Stateless approach.  The number of rows is kept, but the context-aware variables are discarded, reducing the number of columns.}
    \label{tab:stateless}
    \begin{tabular}{c c c c}
        \toprule
        $X^P_1$ & \dots & $X^P_{n_p}$ & $Y$ \\
        \midrule
        \dots & \dots & \dots & \dots \\
        \dots & \dots & \dots & \dots \\
        \dots & \dots & \dots & \dots \\
        \dots & \dots & \dots & \dots \\
        %\dots & \dots & \dots & \dots \\
        \bottomrule
    \end{tabular}
  \end{subtable}%
  \hfill
  \begin{subtable}[t]{0.5\columnwidth}    
    \centering 
    \caption{Stateful approach. The number of rows is reduced since all observations that share the same context ($X^C_1$, \dots, $X^C_{n_c}$) are combined.  New variables ($\Gamma_1$, \dots, $\Gamma_{n_\gamma}$) are the result of the aggregation operations.}
    \label{tab:stateful}
    \begin{tabular}{c c c c c c c}
        \toprule
        $X^C_1$ & \dots & $X^C_{n_c}$ & $\Gamma_1$ & \dots & $\Gamma_{n_\gamma}$ & $Y$ \\
        \midrule
        \dots & \dots & \dots & \dots & \dots & \dots & \dots \\
        \dots & \dots & \dots & \dots & \dots & \dots & \dots \\
        \bottomrule
    \end{tabular}
  \end{subtable}%

\end{table}

\section{Results and Discussion}\label{results}
\begin{wrapfigure}{r}{0.5\textwidth}
    \centering
    \includegraphics[width=0.4\textwidth]{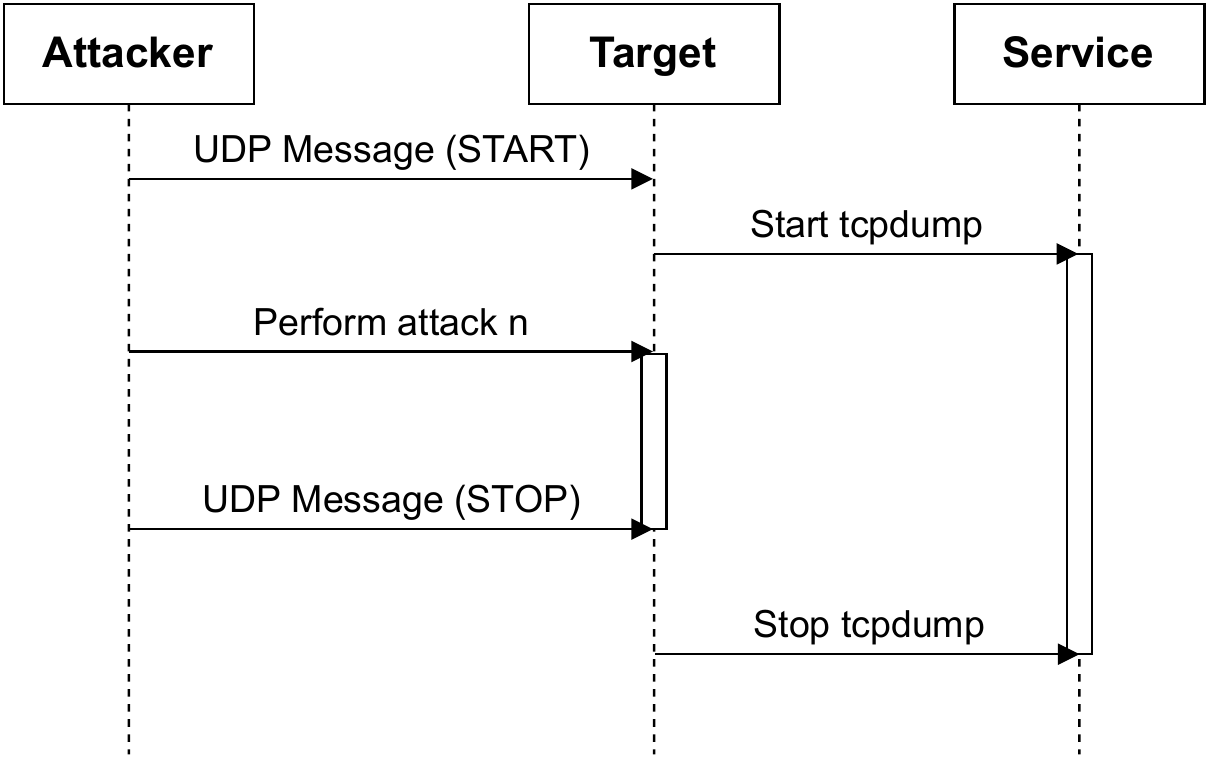}
    \caption{Attack dataset lifeline. Managing the service to capture packets through UDP messages between Attacker and Target.}
    \label{fig:attack_pipeline}
%\end{figure}
\end{wrapfigure}

We created an environment using a cloud service provider to create a geographically distributed test-bed. For the attacker instance, the container's configuration file is available on our public repository. To support the labeling process, we set up a UDP daemon on each of the cloud instances that receive remote UDP commands from the attacker container to start and stop the network traffic capture using \texttt{Tcpdump}\footnote{Tcpdump: tcpdump.org} on these cloud instances. The choice for the UDP protocol is because we focused on the TCP protocol, so there is no conflict between the command and control traffic and the one that makes up our dataset. 
Figure~\ref{fig:attack_pipeline} presents the overall picture of the process. From the UDP message starting the recording, the source IP address (i.e., attacker address) is used to compose the \texttt{Tcpdump} filter to capture only the traffic from the attacker to the target instance. 

%\begin{figure}[h]

Hence, we generated a specific file containing the network traffic with a filename composed by the label name (e.g., attack type) and its start timestamp for each attack. Finally, all network traffic files generated during the attack are retrieved from multiple cloud instances, processed to retrieve the network traffic, and merged into a single dataset.

\begin{figure}[h]
    \centering
    \includegraphics[width=\textwidth]{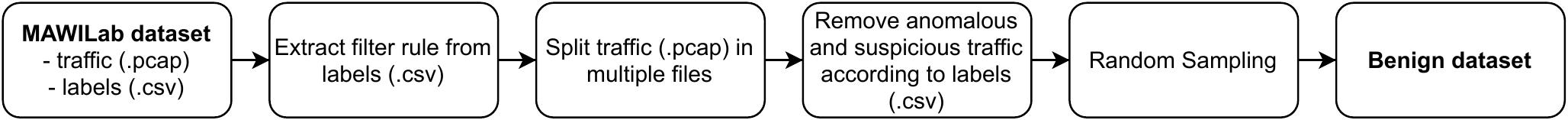}
    \caption{Benign dataset process. After obtaining the MAWILab traffic, and the labels for anomalous/suspicious traffic, the processing is performed to obtain a benign dataset.}
    \label{fig:benign_pipeline}
\end{figure}

For the benign dataset, we use the up-to-date MAWILab dataset, which is a daily 15-minutes network recording from trans-pacific backbone traffic between the USA and Japan. It has labels for the following classes: anomalous and suspicious. Our rationale to obtain benign traffic is to remove from the MAWILab dataset all the traffic inferred as anomalous and suspicious by the MAWILab classifiers. Normally, MAWILab traffic contains tens of millions of packets on each trace, so we introduced a step to random sample packets from these filtered traces to work with a not too much-imbalanced dataset. It is important to highlight that our current applications use a stateless approach (analysis of each packet without the context), so for a stateful analysis (based on flow), the sampling step must retrieve packets that are part of the same context. 
Figure~\ref{fig:benign_pipeline} presents a summary of the process to obtain the benign dataset.

%To reproduce these results, please refer to our public repository presented in Appendix~\ref{appendix}.

% \subsection{Example -- Internet port scanning}
% % Explain the problem
Our research focuses on the problem of the detection of port scanning attacks. This Attack attempts to identify the system's available services or characteristics, which is normally the first step of a cyber attack. Thus, blocking port scanning stops an attack in its early stages, reducing the risks and the resources to secure a system.
%
% % Detail
 We effortlessly generated a dataset for the port scanning problem with this proposed framework for the Internet environment. The dataset comprises $22$ classes (TCP port scanning attacks) targeting $4$ cloud instances, resulting in $455{,}503$ correctly labeled attack samples. For the benign samples, we obtained from MAWILab traffic from November 10\footnote{fukuda-lab.org/mawilab/v1.1/2020/11/10/20201110.html} and 29\footnote{fukuda-lab.org/mawilab/v1.1/2020/11/29/20201129.html}, 2020. After preprocessing the MAWILab traffic to remove the attack samples, we got a total of $380{,}438$ samples (packets) of benign traffic, resulting in a dataset with $835{,}941$ packets with the following distribution: ${Benign}=46\%$ and ${Attack}=54\%$ in a \emph{tidy} dataset with $41$ features (packet-intrinsic and context-aware). Furthermore, the NIDS is an inherently imbalanced problem (higher occurrences of benign traffic); indeed, our solution provides a controlled approach over it, either by managing the repetition of attacks, the number of target instances, or sampling benign traffic. 

\section{Conclusion}
\label{conclusion}
We presented the challenges for ML-based NIDS of aging datasets, the difficulty of coping with the constantly evolving characteristics of network traffic, and bad performance in the real-world environment. To address these gaps, we presented a systematic data-centric approach capable of generating up-to-date NIDS datasets. 
The approach exploits the public and up-to-date MAWILab dataset in conjunction with virtualization to generate the attack traffic. Then, it combines those two sources in a salting process to generate labeled NIDS datasets.
%
%It is a work in progress, and we plan for the next steps of this research to include the generation of features based on network flow as part of the automated dataset generation pipeline. Also, this will enable the performance comparison between models trained on datasets generated by our approach, and the currently available research results obtained with well-established datasets in the current model-centric approach. Moreover, we understand as an open challenge ways to automatic generate up-to-date benign traffic that reliably represent the traffic of envisioned NIDS deployment environments.
The reproducible source-code to generate datasets as presented on this paper is available in our repository: \url{https://github.com/c2dc/AB-TRAP} (steps $A$ and $B$ from \cite{ab-trap}).

\newpage
\bibliographystyle{unsrt}
{
\small
\bibliography{bibfile}

%%%%%%%%%%%%%%%%%%%%%%%%%%%%%%%%%%%%%%%%%%%%%%%%%%%%%%%%%%%%

%%%%%%%%%%%%%%%%%%%%%%%%%%%%%%%%%%%%%%%%%%%%%%%%%%%%%%%%%%%%

%\appendix

%\section{Appendix}\label{appendix}
%The reproducible source-code to generate datasets as presented on this paper is available in our repository: \url{https://github.com/c2dc/AB-TRAP} (steps $1$ and $2$).
}

%Optionally include extra information (complete proofs, additional experiments and plots) in the appendix.
%This section will often be part of the supplemental material.

\end{document}